\documentclass[10pt, journal, twocolumn]{IEEEtran}

\usepackage{color,soul}

\usepackage{amsfonts}
\usepackage{cite}
\usepackage{graphicx}
\usepackage{psfrag}
\usepackage{amsmath}
\usepackage{times}
\usepackage[dvips]{epsfig}
\usepackage{amssymb}
\usepackage{subfigure}
\usepackage[center,small]{caption2}
\usepackage[below]{placeins}

\usepackage{url}
\usepackage{stfloats}

\usepackage{array}
\begin{document}

\title{Alternative Awaiting and Broadcast for Two-Way Relay Fading Channels}

\author{Jianquan Liu, Meixia Tao, \emph{Senior Member, IEEE}, and Youyun Xu, \emph{Senior Member, IEEE}
\thanks{Copyright (c) 2013 IEEE. Personal use of this material is permitted. However, permission to use this material for any other purposes must be obtained from the IEEE by sending a request to pubs-permissions@ieee.org.}
\thanks{The authors are with the Department of Electronic Engineering, Shanghai Jiao Tong University, Shanghai, P. R. China, 200240.
Youyun Xu is also with the Institute of Communications Engineering, PLA University of Science \& Technology, Nanjing,
P. R. China, 210007. Email: jianquanliu@sjtu.edu.cn, jianquanliu@gmail.com, mxtao@sjtu.edu.cn, yyxu@vip.sina.com.}
}

\maketitle

\IEEEpeerreviewmaketitle

\begin{abstract}

We investigate a two-way relay (TWR) fading channel where two source nodes wish to exchange information with the help of a relay node. Given traditional TWR protocols, transmission rates in both directions are known to be limited by the hop with lower capacity, i.e., the $\min$ operations between uplink and downlink. In this paper, we propose a new transmission protocol, named as \emph{alternative awaiting and broadcast} (AAB), to cancel the $\min$ operations in the TWR fading channels. The operational principles, new upper bound on ergodic sum-capacity (ESC) and convergence behavior of average delay of signal transmission (ST) (in relay buffer) for the proposed AAB protocol are analyzed. Moreover, we propose a suboptimal encoding/decoding solution for the AAB protocol and derive an achievable ergodic sum-rate (ESR) with corresponding average delay of ST. Numerical results show that 1) the proposed AAB protocol significantly improves the achievable ESR compared to the traditional TWR protocols, 2) considering the average delay of system service (SS) (in source buffer), the average delay of ST induced by the proposed AAB protocol is very small and negligible.

\end{abstract}



\section{Introduction}\label{sec:introduction}

Two-way relaying has recently obtained lots of research interests~\cite{JSAC07:Rankov,TON08:Katti,IT08:Kim,AMC06:Zhang,ASC07:Katti}. The classic two-way relay (TWR) channel consists of three nodes, wherein two source nodes exchange information with the help of a relay node. Upon receiving the bidirectional information flows, the relay node combines them together and then broadcasts to the two desired destinations. A number of TWR protocols have been proposed. Among them, four popular protocols are known as amplify-and-forward (AF)~\cite{JSAC07:Rankov,AMC06:Zhang}, decode-and-forward (DF)~\cite{JSAC07:Rankov,ICC07:Popovski}, denoise-and-forward (DNF)~\cite{ICC07:Popovski,IT08:Kim} and compress-and-forward (CF)~\cite{arxiv09:Kim} respectively. Meanwhile, the operation at the relay node resembles network coding\cite{IT00:Ahlswede}. It is often referred to as physical layer network coding (PLNC)~\cite{AMC06:Zhang}.
Based on these transmission protocols, some PLNC methods have also been proposed and analyzed, such as bit-level Exclusive OR (XOR)~\cite{ICC07:Popovski}, symbol-level superposition~\cite{JSAC07:Rankov}, bit/symbol-level superimposed XOR~\cite{GC09:We} and codeword-level modulo addition~\cite{IT10:Wilson,IT10:Nam}.

However, all the aforementioned TWR protocols carry out immediate forwarding (i.e., no buffer) on the whole of received information flows at the relay, so both the transmission rates in two opposite directions are known to be limited by the hop with lower capacity, i.e., the $\min$ operation between the uplink and downlink in an identical direction. As shown in Fig.~\ref{fig:system_model}, we have $R_{02}[t]=\min(R_{01}[t],R_{12}[t])$ and $R_{20}[t]=\min(R_{21}[t],R_{10}[t])$, where $R_{ij}[t]$ denote the transmission rate of the link from node $i$ to node $j$ during the $t^{th}$ round of information exchange (also denoted as the $t^{th}$ time unit), for $i,j \in \{0,1,2\}$. It may be unavoidable in the TWR Gaussian channels because the channel gain of the same link is stationary during the former and the latter time units, e.g., $h_{12}[t_1]=h_{12}[t_2]$. Fortunately, due to quick variation of channel gains (e.g., $h_{12}[t_1] \neq h_{12}[t_2]$), the TWR fading channels have the potential to eliminate the $\min$ operations by introducing certain delay of partial information exchange. To the best of our knowledge, none of the works in the literatures has considered the former problem and sufficiently exploited the potential benefits of asymmetric channel gains for the TWR fading channels.

In this paper, we propose a new transmission protocol, named as \emph{alternative awaiting and broadcast} (AAB), to eliminate the $\min$ operations between the uplink and downlink in an identical direction in the TWR fading channels. The operational principles, new upper bound on ergodic sum-capacity (ESC) and convergence behavior of average delay of signal transmission (ST) (in relay buffer) for the proposed AAB protocol are analyzed. Moreover, we propose a suboptimal encoding/decoding solution for the AAB protocol and derive an achievable ergodic sum-rate (ESR) with corresponding average delay of ST. Numerical results show that 1) the proposed AAB protocol significantly improves the achievable ESR compared to the traditional TWR protocols, 2) considering the average delay of system service (SS) (in source buffer), the average delay of ST induced by the proposed AAB protocol is very small and negligible.

\section{System Model}\label{sec:system_model}
\begin{figure}
\centering
\includegraphics[width=3in]{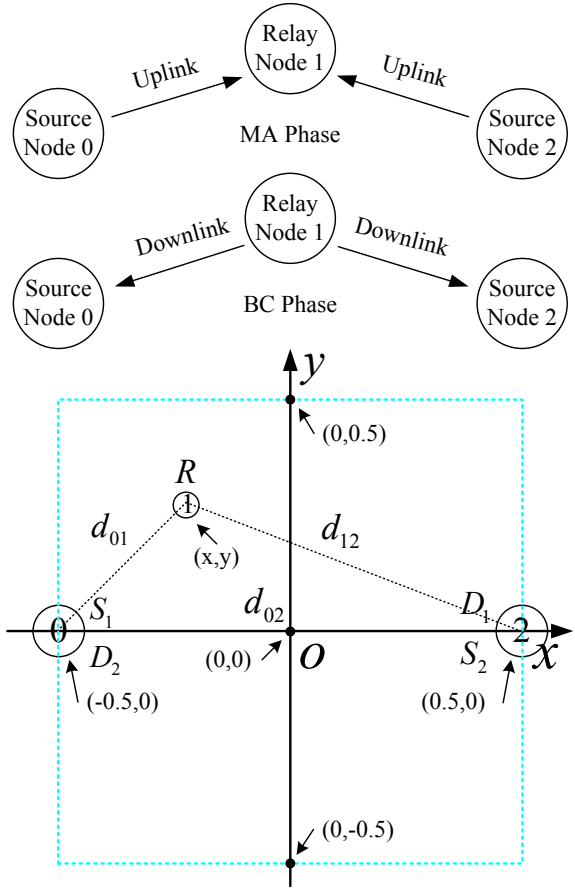}
\caption{System model of two-way relay (TWR) fading channels.}
 \label{fig:system_model}
\end{figure}

We consider a classic three-node TWR fading channel as shown in Fig.~\ref{fig:system_model}, where two source nodes, denoted as $0$ and $2$, wish to exchange information with the help of a relay node, denoted as $1$. The channel on each communication link is assumed to be corrupted with fading and additive white Gaussian noise (AWGN).
Let $t$ denote the $t^{th}$ round of source information exchange (SIE), for all $t$. The instantaneous SNR from node $i$ to node $j$ in the $t^{th}$ round is denoted as $\gamma_{ij}[t]=\frac{P_i|h_{ij}[t]|^2}{\sigma_j^2}$, for $i,j\in\{0,1,2\}$. It counts the $t^{th}$ channel gain $h_{ij}[t]$ from node $i$ to node $j$, average transmit power $P_i$ at the node $i$ and AWGN power $\sigma_j^2$ at the node $j$. Note that $|\cdot|$ stands for the magnitude of a complex scalar. The ergodic capacity $\bar{C}_{ij}$ in $bit/s/Hz$ is determined as $\bar{C}_{ij} = {\cal E}\Big\{C_{ij}[t]\Big\} = {\cal E}\Big\{C(\gamma_{ij}[t])\Big\} = {\cal E}\Big\{\log_2(1+\gamma_{ij}[t])\Big\}$, where ${\cal E}\{\cdot\}$ represents the expectation operator. For simplicity, we assume the channel gains are reciprocal and unchanged during one round of SIE. Then, we have $h_{01}[t]=h_{10}[t]$, $h_{21}[t]=h_{12}[t]$. We also assume that $h_{01}[t]$ and $h_{21}[t]$ are mutually independent and subject to an identical distribution, for all $t$. That is to say, $h_{01}$ and $h_{21}$ are independent and identically distributed (i.i.d.).

In this paper, we focus on two-phase TWR system with equal time slot, which can be divided into a multiple access (MA) phase and a broadcast (BC) phase, as depicted in Fig.~\ref{fig:system_model}. In the MA phase, two source nodes transmit simultaneously and the relay node listens. Two MA channels are also denoted as two uplinks. In the BC phase, the relay node transmits while two source nodes listen. We also denote two BC channels as two downlinks. We assume that all the nodes operate in the half-duplex mode. Let $P_0=P_1=P_2=P, \sigma_0^2=\sigma_1^2=\sigma_2^2=\sigma^2$. We use bold upper letters to denote vectors and lower letters to denote elements. We introduce an \emph{ergodic sum-rate} (ESR) to describe the performance of the TWR fading channels. An instantaneous sum-rate of $R_s[t]$ is said to be achievable if, there exist at least an encoding/decoding scheme of rate $R_0[t], R_2[t], R_0[t] + R_2[t] \leq R_s[t]$ for two source nodes respectively, with as small probability of instantaneous error in the $t^{th}$ round of SIE as desired. An ESR of $\bar{R}_s$ is considered as the average sum-rate over all channel distributions, i.e., $\bar{R}_s = {\cal E}\{R_s\}$. The instantaneous sum-capacity $C_s$ and \emph{ergodic sum-capacity} (ESC) $\bar{C}_s$ are then the supremums of $R_s$ and $\bar{R}_s$, respectively.

For ease of understanding, we expound four important definitions before introducing our work.

a) Round of source information exchange (SIE): One round of SIE is defined as one round of information exchange. In one round of SIE, the received information at each destination node may not include all the information transmitted from the corresponding source node.

b) Round of desired information exchange (DIE): One round of DIE maybe contain several rounds of SIE. In one round of DIE, the received information at each destination node includes all the information transmitted in the first round of SIE from the corresponding source node.

c) Delay of signal transmission (ST): In this delay, we do not care at each source node whether the source information waits to be transmitted or not. We just consider that how many extra time units (begin with the second round of SIE) do we need to complete one round of DIE.

d) Delay of system service (SS): In this delay, we consider that at each source node how many extra time units (begin with the second round of SIE) does the source information spend waiting to be transmitted.

\section{Proposed Alternative Awaiting and Broadcast (AAB)}\label{sec:AAB}

\subsection{Operational principles}\label{sec:AAB_OP}
\begin{figure*}[t]
\centering
\includegraphics[width=6in]{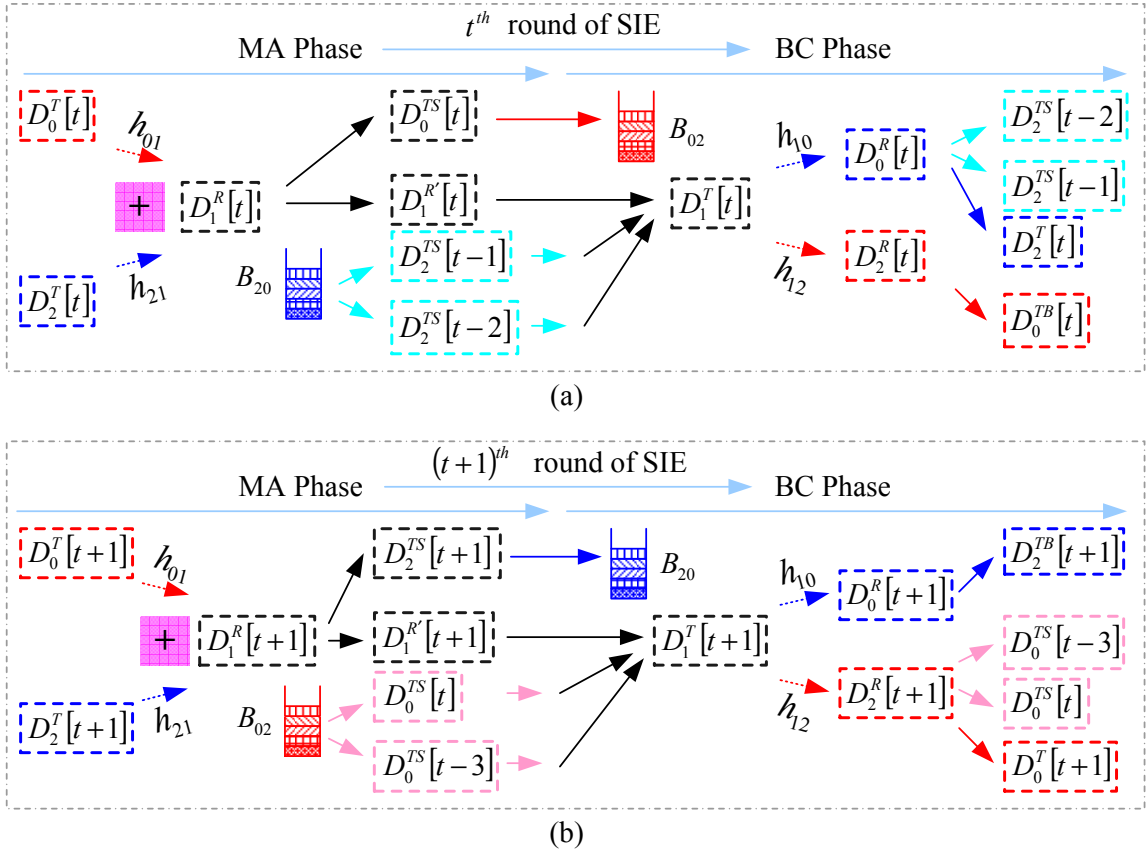}
\caption{Operational principles of proposed alternative awaiting and broadcast (AAB) protocol.}
 \label{fig:AAB_OP}
\end{figure*}

Let $t$ denote the $t^{th}$ round of SIE, for all $t$. We assume that each round of SIE occupies one time unit.
Then, each round of SIE is divided into two time slots. One is for a MA phase and the other is for a BC phase.
We first summarize some important notations as follows:
$D_{i}^T[t]$: the information packet to be transmitted from the node $i$ during the $t^{th}$ round of SIE, for $i \in \{0,1,2\}$.
$D_{i}^{TB}[t]$: the sub-packet split from the transmitted packet $D_{i}^T[t]$ and to be broadcasted immediately, for $i \in \{0,2\}$.
$D_{i}^{TS}[t]$: the sub-packet split from the transmitted packet $D_{i}^T[t]$ and to be stored for delayed transmission, for $i \in \{0,2\}$.
$D_{i}^R[t]$: the information packet to be received at the node $i$, during the $t^{th}$ round of SIE, for $i \in \{0,1,2\}$.
$|D_{i}[t]|$: the length in bits of the packet $D_{i}[t]$ during the $t^{th}$ round of SIE.
$B_{ij}$: the relay buffer which is set only for the transmission from node $i$ to node $j$, for $\{i,j\} \in \{0,2\}$.

Without loss of generality, we assume that we have
$|h_{01}[t-3]|^2 \geq |h_{21}[t-3]|^2$,
$|h_{01}[t-2]|^2 \leq |h_{21}[t-2]|^2$,
$|h_{01}[t-1]|^2 \leq |h_{21}[t-1]|^2$,
$|h_{01}[t]|^2 \geq |h_{21}[t]|^2$,
and $|h_{01}[t+1]|^2 \leq |h_{21}[t+1]|^2$ in a TWR system with reciprocal fading channels.

\subsubsection{$t^{th}$ round of SIE --- a MA phase and a BC phase}\label{sec:AAB_SIE_t}

As depicted in Fig.~\ref{fig:AAB_OP}-(a), we firstly consider a MA phase. Wherein, two source nodes transmit their information packet, $D_0^T[t]$ and $D_2^T[t]$, to the relay node simultaneously. Since $|h_{01}[t]|^2 \geq |h_{21}[t]|^2$, we have $|D_0^T[t]| \geq |D_2^T[t]|$. Due to the application of PLNC, the information packet $D_1^R[t]$ received at the relay node is a function of two transmitted packets, i.e., $D_1^R[t]=\bold{F}\Big(D_0^T[t],D_2^T[t]\Big)$.

According to the capacity region\footnote{The capacity region in the BC phase~\cite{IT08:Oechtering} is given as $\{R_{10}[t],R_{12}[t]\}:R_{10}[t] \leq C_{10}[t], R_{12}[t] \leq C_{12}[t]$. Due to $|h_{01}[t]|^2 \geq |h_{21}[t]|^2$ and $h_{01}[t]=h_{10}[t], h_{21}[t]=h_{12}[t]$, we have $C_{12}[t] \leq C_{10}[t]$. With $|D_0^T[t]|=C_{01}[t]=C_{10}[t]$, we gain $|D_0^T[t]| \geq C_{12}[t]$. This is the reason why $D_0^T[t]$ can not be decoded successfully. Likewise, the source node 0 can receive more than $D_2^T[t]$.} in the BC phase~\cite{IT08:Oechtering} and $h_{10}[t] = h_{01}[t]$, $h_{12}[t] = h_{21}[t]$, the desired information packet $D_0^T[t]$ can not be decoded successfully at the source node 2 if we broadcast $D_1^R[t]$ directly. At the same time interval, the source node 0 can receive more information than the transmitted packet $D_2^T[t]$. Thus, we propose a new packet processing method which is denoted as extracting and embedding (EDE). This new method operates on the received packet at the relay node and is elaborated as follows:

a) Due to $|h_{10}[t]|^2 \geq |h_{12}[t]|^2$, the relay node extracts a sub-packet $D_0^{TS}[t]$ from the received packet $D_1^R[t]$ under the rule of $|D_0^T[t]|-|D_0^{TS}[t]| \leq C_{12}[t]$. Now the received packet $D_1^R[t]$ is changed into a distinct packet $D_1^{R'}[t]$ with $D_1^{R'}[t]=\bold{F}\Big(D_0^{TB}[t],D_2^T[t]\Big)$. Therein, the sub-packet $D_0^{TB}[t]$ is remainder information of the transmitted packet $D_0^T[t]$ after extracting a sub-packet $D_0^{TS}[t]$. We have $|D_0^{TB}[t]|=|D_0^T[t]|-|D_0^{TS}[t]|$. Then we store $D_0^{TS}[t]$ in the relay buffer $B_{02}$.

b) Similar to the analysis in method a), due to $|h_{01}[t-2]|^2 \leq |h_{21}[t-2]|^2$, $|h_{01}[t-1]|^2 \leq |h_{21}[t-1]|^2$, $|h_{10}[t-2]|^2 \leq |h_{12}[t-2]|^2$ and $|h_{10}[t-1]|^2 \leq |h_{12}[t-1]|^2$, two sub-packets $D_2^{TS}[t-1]$ and $D_2^{TS}[t-2]$ had been stored in the relay buffer $B_{20}$ during the $(t-1)^{th}$ and $(t-2)^{th}$ rounds of SIE respectively. If we have $|D_2^{TS}[t-2]|+|D_2^{TS}[t-1]|+|D_2^T[t]| \leq C_{10}[t]$, the aforementioned two sub-packets will be picked up and embedded into the packet $D_1^{R'}[t]$. Through the secondly application of the PLNC, we generate the broadcasted packet $D_1^T[t]$ which is a function of three sub-packets (i.e., $D_0^{TB}[t]$, $D_2^{TS}[t-1]$ and $D_2^{TS}[t-2]$) and an original packet $D_2^T[t]$.

The storage and extraction in each relay buffer obey the rule of First-In First-Out (FIFO).
Then, the relay node broadcasts the generated packet $D_1^T[t]$ during the coming BC phase.
With the help of its self-information, each destination node can successfully decode the exchanged information transmitted by the corresponding source node. Based on the known information packets, such as $D_0^T[t-2],D_0^T[t-1]$ and $D_0^T[t]$, the destination node 0 can easily decode two sub-packets (i.e., $D_2^{TS}[t-2]$ and $D_2^{TS}[t-1]$) and an original packet $D_2^T[t]$, from the received packet $D_0^R[t]$. At the same time, the node 2 can also successfully decode the only one sub-packet, e.g., $D_0^{TB}[t]$, from the received packet $D_2^R[t]$ with the help of a known packet $D_2^T[t]$. Wherein, we have $D_0^R[t]=D_2^R[t]=D_1^T[t]$.

\subsubsection{$(t+1)^{th}$ round of SIE --- a MA phase and a BC phase}\label{sec:AAB_SIE_t_1}

Due to that the main difference from the previous subsection is the channel condition only, the remaining operations can be done in the same manner as depicted in Fig.~\ref{fig:AAB_OP}-(b).

\subsubsection{$t^{th}$ round of DIE}\label{sec:AAB_DIE_t}

Two transmitted packets, $D_0^T[t]$ and $D_2^T[t]$, are successfully exchanged between two source nodes during the former two rounds of SIE, so the $t^{th}$ round of DIE contains the $t^{th}$ and $(t+1)^{th}$ rounds of SIE. Here, we can say that the delay of ST of the $t^{th}$ round of DIE is one time unit.

\subsection{New upper bound on ergodic sum-capacity (ESC)}\label{sec:AAB_AESR}

According to the upper bound on capacity in the MA phase~\cite{IT08:Kim} and the exact capacity in the BC phase~\cite{IT08:Oechtering}, we obtain the upper bound on ESC of the traditional TWR protocols given as
\begin{eqnarray}\label{eq:UB_sum_capacity_T}
\bar{C}_s^{uT} &=& \frac{1}{2}{\cal E} \bigg\{\min\Big\{C_{01}[t],C_{12}[t]\Big\} + \min\Big\{C_{21}[t],C_{10}[t]\Big\} \bigg\} \nonumber \\
&=& {\cal E} \bigg\{ \min\Big\{C_{01}[t],C_{21}[t]\Big\} \bigg\}.
\end{eqnarray}

If we introduce the AAB protocol, the message received at the relay node with a higher rate may be broadcasted in the $t^{th}$ and some successive $t'^{th}$ rounds of SIE, where $t' > t$. At the same time, the relay node can also broadcast both the message received with a lower rate and some accumulated message during the former $t''^{th}$ rounds of SIE, where $t'' < t$. Then, a new upper bound on the ESC is obtained as follow.
\begin{equation}\label{eq:UB_sum_capacity}
\bar{C}_s^u =  {\cal E} \Big\{\frac{1}{2}C_{01}[t] + \frac{1}{2}C_{21}[t] \Big\} = \frac{1}{2} \Big(\bar{C}_{01}+ \bar{C}_{21} \Big).
\end{equation}

We can see that (\ref{eq:UB_sum_capacity_T}) suffers from the $\min$ operations while the proposed AAB protocol removes it from a new upper bound on ESC in (\ref{eq:UB_sum_capacity}).

\subsection{Convergence Behavior of Average Delay of ST}\label{sec:AAB_ADRF}

Let $R_{01}[t]-R_{21}[t] = \theta(C_{01}[t]-C_{21}[t]) = \theta(C_{10}[t]-C_{12}[t])$. Here we regard $\theta$ as 1) the ratio of the difference of two transmission rates obtained from the capacity region to that from the upper bound in the MA phase, 2) the ratio of the difference of two transmission rates obtained in the MA phase to that obtained in the BC phase. Without loss of generality, we assume that $|h_{01}[t]|^2 \geq |h_{21}[t]|^2$ and $R_{01}[t]-R_{21}[t]$ can be successfully broadcasted during $l_{01}[t]$ time units --- from the $(t+1)^{th}$ to the $(t+l_{01}[t])^{th}$ time unit. Let $\Gamma[t] = (\frac{\sigma^2+P|h_{01}[t]|^2}{\sigma^2+P|h_{21}[t]|^2})^\theta$ and $\Delta\Big[t+l(t)\Big] = \frac{\sigma^2+P|h_{12}[t+l(t)]|^2}{\sigma^2+P|h_{10}[t+l(t)]|^2}$, we obtain
\begin{eqnarray}
\log_2\Big(\Gamma[t]\Big) &\leq& \sum\limits_{l(t)=l_{01}'[t]+1}^{l_{01}[t]}\phi\Big(l(t)\Big)\log_2\Big(\Delta\Big[t+l(t)\Big]\Big) \nonumber\\
&&+ \sum\limits_{l(t_0)=l_{01}'[t_0]+1}^{l_{01}[t_0]}\phi\Big(l(t_0)\Big)\log_2\Big(\Delta\Big[t_0+l(t_0)\Big]\Big)\nonumber\\
&&- \log_2\Big(\Gamma[t_0]\Big),
\end{eqnarray}
where $t_0$ denotes the time unit in which adjacent former information $R_{01}[t_0]-R_{21}[t_0]$, namely $\log_2\Big(\Gamma[t_0]\Big)$, has been generated. Moreover, $l_{01}'[t]=l_{01}[t_0]-(t-t_0)$ denotes the extra time units should be used from the current time unit $t$ in order to successfully broadcast the adjacent former information $\log_2\Big(\Gamma[t_0]\Big)$. Note that $\phi(l(t))$ satisfies $\phi(l(t))= 1$ only if $|h_{01}[t+l(t)]|^2 \leq |h_{21}[t+l(t)]|^2$, for $l(t)\in \Big(l_{01}'[t], l_{01}[t]\Big]$, else $\phi(l(t))= 0$. Then we obtain an instantaneous delay of ST $l_{01}[t]$ given as
\begin{eqnarray}\label{eq:ad}
l_{01}[t] &=& \min\limits_{l_{01}[t] \in {\cal Z}^+}\Bigg\{ \prod\limits_{l(t)=l_{01}'[t]+1}^{l_{01}[t]}\Big(\Delta\Big[t+l(t)\Big]\Big)^{\phi(l(t))}\nonumber\\
&& \times \prod\limits_{l(t_0)=l_{01}'[t_0]+1}^{l_{01}[t_0]}\Big(\Delta\Big[t_0+l(t_0)\Big]\Big)^{\phi(l(t_0))}\nonumber\\
&&\geq \Gamma[t]\times\Gamma[t_0]\Bigg\}.
\end{eqnarray}

Similarly, the instantaneous delay of ST $l_{21}[t]$ is obtained easily when $|h_{01}[t]|^2 \leq |h_{21}[t]|^2$. The average delay of ST can be written as ${\cal L}=\Big\{{\cal E}\{l_{01}[t]\},{\cal E}\{l_{21}[t]\}\Big\}$.

\section{Achievable Ergodic Sum-Rate}\label{sec:AESR}

\subsection{A Suboptimal Encoding and Decoding Solution}\label{sec:AESR_EDS}
\begin{figure*}[t]
\centering
\includegraphics[width=6.5in]{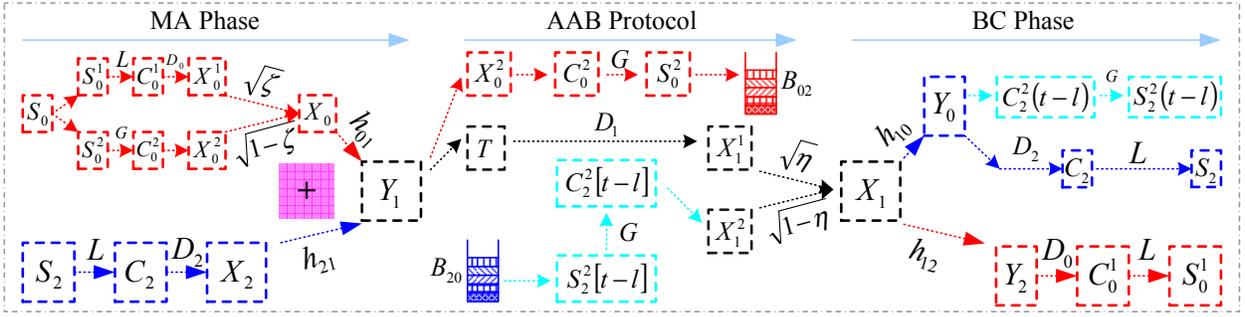}
\caption{A suboptimal encoding/decoding solution for AAB protocol.}
 \label{fig:AAB_suboptimal}
\end{figure*}

Firstly, we consider a MA phase in the $t^{th}$ round of SIE. Without loss of generality, we assume that $|h_{01}[t]|^2 \geq |h_{21}[t]|^2$, namely $R_{01}[t] \geq R_{21}[t]$. As depicted in Fig.~\ref{fig:AAB_suboptimal} (Left-hand), the source node 0 splits the message $\bold{S}_0[t]$ into two parts: $\bold{S}_0^1[t]$ and $\bold{S}_0^2[t]$. Therein, the length of one part, e.g., $\bold{S}_0^1[t]$, is equal to that of the message $\bold{S}_2[t]$ from the source node 2. $\bold{S}_0^1[t]$ and $\bold{S}_0^2[t]$ are then encoded to $\bold{C}_0^1[t]$ and $\bold{C}_0^2[t]$ by a Lattice code $\bold{L}$ and a Gaussian code $\bold{G}$ respectively. After operating $\bold{X}_0^1[t]=(\bold{C}_0^1[t]+\bold{D}_0[t])\mod\bigwedge^n$ and modulating $\bold{C}_0^2[t]$ to $\bold{X}_0^2[t]$, the source node 0 forms the transmitted signal $\bold{X}_0[t]=\sqrt{\zeta[t]}\bold{X}_0^1[t]+\sqrt{1-\zeta[t]}\bold{X}_0^2[t]$ by superposition and a power allocation coefficient $\zeta[t]$. At the same time, the source node 2 generates the transmitted signal $\bold{X}_2[t]$ through mapping the message $\bold{S}_2[t]$ to $\bold{C}_2[t]$ by an identical Lattice code $\bold{L}$ and operating $\bold{X}_2[t]=(\bold{C}_2[t]+\bold{D}_2[t])\mod\bigwedge^n$. The random dither vectors $\bold{D}_0[t]$ and $\bold{D}_2[t]$ are mutually independent of each other and are also known at both the relay node and two source nodes.

At the relay node, the received superimposed signal is given as
\begin{eqnarray}
\bold{Y}_1[t] &=& h_{01}[t]\bold{X}_0[t]+h_{21}[t]\bold{X}_2[t]+\bold{Z}_1[t]\\
&=& h_{01}[t]\Big(\sqrt{\zeta[t]}\bold{X}_0^1[t]+\sqrt{1-\zeta[t]}\bold{X}_0^2[t]\Big)\nonumber \\
&&+ h_{21}[t]\bold{X}_2[t]+\bold{Z}_1[t]\\
&=& \bold{T}[t]+\sqrt{1-\zeta[t]}h_{01}[t]\bold{X}_0^2[t]+\bold{Z}_1[t].
\end{eqnarray}
As shown in Fig.~\ref{fig:AAB_suboptimal} (Middle), the relay node first decodes $\bold{S}_0^2[t]$ ($\bold{X}_0^2[t]$) by treating a function $\bold{T}[t]$ as noise. Subtracting $\bold{X}_0^2[t]$ off its received signals, then $\bold{T}[t]$ is decoded. Obviously, we should set $\zeta[t]=\frac{|h_{21}[t]|^2}{|h_{01}[t]|^2}$, $\zeta[t] \in [0,1]$, in order to satisfy that two Lattice coded signals have the same received SNR, i.e., $\gamma_{01}^1[t]=\gamma_{21}[t]$. Then, the relay operates $(\bold{T}[t]+\bold{D}_1[t])\mod\bigwedge^n$ and forms $\bold{X}_1^1[t]$ by a random dither vector $\bold{D}_1[t]$. Due to the reciprocity between two relay channels, we have $|h_{12}[t]|^2 \leq |h_{10}[t]|^2$, namely $R_{12}[t] \leq R_{10}[t]$. Therefore, the relay stores $\bold{S}_0^2[t]$ and waits a favorable channel gain, e.g., $|h_{12}[t]|^2 \geq |h_{10}[t]|^2$, to broadcast $\bold{S}_0^2[t]$. At the same time, a fractional message, e.g., $\bold{S}_2^2[t-l]$, for $l \in {\cal Z}^+$, which has been received and stored in the former $(t-l)^{th}$ round of SIE at the relay node, may be picked up and encoded to form $\bold{C}_2^2[t]$ by a Gaussian code. $\bold{C}_2^2[t]$ will be modulated to $\bold{X}_1^2[t]$ and superimposed with $\bold{X}_1^1[t]$ for generating the transmitted signal $\bold{X}_1[t]=\sqrt{\eta[t]}\bold{X}_1^1[t]+\sqrt{1-\eta[t]}\bold{X}_1^2[t]$. Here, $\eta[t]$ is also a power allocation coefficient with $\eta[t] \in [0,1]$. The received Gaussian coded message at the relay node is decoded and stored in buffer $B_{02}$. It will be transmitted during certain rounds of SIE after the current $t^{th}$ round of SIE while the Lattice coded message $\bold{T}[t]$ will be transmitted immediately in the current $t^{th}$ round of SIE. Note that the storage and extraction of each received Gaussian coded message obey the rule of First-In First-Out (FIFO). For example, $\bold{S}_i^2[t_1-l]$ should be broadcasted earlier than $\bold{S}_i^2[t_2-l]$, if we have $t_2 > t_1$ for $\{t_1, t_2\} \in {\cal N}, l \in {\cal Z}^+, i \in \{0,2\}$.

In the BC phase, the superimposed signal $\bold{X}_1[t]$ is broadcasted to two source nodes by the relay node, as depicted in Fig.~\ref{fig:AAB_suboptimal} (Right-hand). At two source nodes, the received signals are given as
$
\bold{Y}_i[t] = h_{1i}[t]\bold{X}_1[t] + \bold{Z}_i[t]= h_{1i}[t]\Big( \sqrt{\eta[t]}\bold{X}_1^1[t]+\sqrt{1-\eta[t]}\bold{X}_1^2[t]\Big)+ \bold{Z}_i[t],
$
where $i \in \{0,2\}$. Since that $\bold{S}_2^2[t-l]$ is known, the source node 2 first subtracts $\bold{X}_1^2[t]$ off its received signal and then decodes the Lattice coded message $\bold{S}_0^1[t]$ by using a Lattice code book $\Big\{\bold{T}, \bold{C}_0^1 \in \bigwedge^n\Big\}$. Meanwhile, the source node 0 first decodes the Lattice coded message, $\bold{S}_2[t]$ ($\bold{X}_1^1[t]$), by using a Lattice code book $\Big\{\bold{T}, \bold{C}_2 \in \bigwedge^n\Big\}$ and treating $\bold{X}_1^2[t]$ as noise. Subtracting $\bold{X}_1^1[t]$ off its received signal, the source node 0 then decodes the Gaussian coded message $\bold{S}_2^2[t-l]$ from $\bold{X}_1^2[t]$ successfully.

\subsection{Achievable Ergodic Sum-Rate (ESR)}\label{sec:AESR_AESR}

\subsubsection{Achievable ergodic sum-rate (ESR)}\label{sec:AESR_AESR_AESR}

Let $R_{ij}[t], i,j \in \{0,2\}$, denote the instantaneous transmission rate of one side with transmission direction $i \rightarrow j$ during the $t^{th}$ round of SIE. According to the analysis in Subsection~\ref{sec:AESR_EDS}, we obtain an instantaneous rate pair $(R_{01}[t],R_{21}[t])$ in the MA phase given by
\begin{eqnarray}\label{eq:AESR_AESR_6}
R_{01}[t] &\leq& \frac{1}{2}\Big\{\Big[\log_2(\frac{1}{2}+\frac{P|h_{21}[t]|^2}{\sigma^2})\Big]^+ \nonumber\\
          && + \log_2(1+\frac{P(|h_{01}[t]|^2-|h_{21}[t]|^2)}{\sigma^2+2P|h_{21}[t]|^2})\Big\},\\
\label{eq:AESR_AESR_7}
R_{21}[t] &\leq& \frac{1}{2}\Big[\log_2(\frac{1}{2}+\frac{P|h_{21}[t]|^2}{\sigma^2})\Big]^+,
\end{eqnarray}
and an instantaneous rate pair $(R_{10}[t],R_{12}[t])$ in the BC phase as
\begin{eqnarray}\label{eq:AESR_AESR_8}
R_{12}[t] &\leq& \frac{1}{2}\log_2(1+\frac{\eta[t] P|h_{12}[t]|^2}{\sigma^2}),\\
\label{eq:AESR_AESR_9}
R_{10}[t] &\leq& \frac{1}{2}\Big\{\log_2(1+\frac{\eta[t] P|h_{10}[t]|^2}{\sigma^2+(1-\eta[t]) P|h_{10}[t]|^2})\nonumber\\ 
&&+ \log_2(1+\frac{(1-\eta[t])P|h_{10}[t]|^2}{\sigma^2})\Big\},
\end{eqnarray}
where $\eta[t] \in [0,1]$.

Then, we achieve an instantaneous rate pair, denoted as $(R_{02}[t],R_{20}[t])$, given by
\begin{eqnarray}
R_{02}[t] \leq \max\Big\{R_{01}[t],R_{12}[t]\Big\}, \nonumber\\
R_{20}[t] \leq \max\Big\{R_{21}[t],R_{10}[t]\Big\}.
\end{eqnarray}

In general, an achievable ESR of the proposed AAB protocol is given by
\begin{eqnarray}\label{eq:AESR_AESR_11}
\bar{R}_s^{AAB} \leq {\cal E}\Bigg\{\bigg[\log_2\Big(\frac{1}{2}+\frac{P\min\{|h_{01}[t]|^2,|h_{21}[t]|^2\}}{\sigma^2}\Big)\bigg]^+ \nonumber \\
+ \frac{1}{2} \log_2\bigg(1+\frac{P\Big||h_{01}[t]|^2-|h_{21}[t]|^2\Big|}{\sigma^2+2P\min\Big\{|h_{01}[t]|^2,|h_{21}[t]|^2\Big\}}\bigg)\Bigg\}.
\end{eqnarray}

For comparisons, we also give an achievable ESR of the DNF protocol given as~\cite{IT10:Wilson}
\begin{eqnarray}\label{eq:AESR_DNF}
\bar{R}_s^{DNF} \leq {\cal E}\left\{\left[\log_2\left(\frac{1}{2}+\frac{P\min\{|h_{01}[t]|^2,|h_{21}[t]|^2\}}{\sigma^2}\right)\right]^+ \right\}.
\end{eqnarray}

We can see that the proposed AAB protocol achieves an extra improvement on ESR.

\subsubsection{Power allocation at relay node}\label{sec:AESR_AESR_PAR}

Satisfying that we should not decrease the transmission rate of the side with inferior uplink, e.g., source node 2, we need
\begin{eqnarray}\label{eq:AESR_AESR_12}
\frac{1}{2} \log_2\Big(1+\frac{\eta[t] P|h_{12}[t]|^2}{\sigma^2}\Big) 
\geq \frac{1}{2}\log_2\Big(\frac{1}{2}+\frac{P|h_{21}[t]|^2}{\sigma^2}\Big), \\
\frac{1}{2}\log_2\Big(1+\frac{\eta[t] P|h_{10}[t]|^2}{\sigma^2+(1-\eta[t])P|h_{10}[t]|^2}\Big) \nonumber\\
\geq \frac{1}{2}\log_2\Big(\frac{1}{2}+\frac{P|h_{21}[t]|^2}{\sigma^2}\Big).
\end{eqnarray}
After simplifications, we obtain
\begin{equation}\label{eq:AESR_AESR_16}
\eta[t] \geq\left\{ \begin{array}{clcc}
1- \frac{\sigma^2}{2P|h_{21}[t]|^2}, &{\text{if}}& 0 < \frac{|h_{21}[t]|^2}{|h_{10}[t]|^2} \leq \frac{1}{2},  \\
&&\frac{P|h_{21}[t]|^2}{\sigma^2} \geq \frac{1}{2},\\
\frac{(2P|h_{21}[t]|^2-\sigma^2)(P|h_{10}[t]|^2+\sigma^2)}{P|h_{10}[t]|^2(\sigma^2+2P|h_{21}[t]|^2)}, &{\text{if}}& \frac{1}{2} < \frac{|h_{21}[t]|^2}{|h_{10}[t]|^2} \leq 1,\\
&& \frac{P|h_{21}[t]|^2}{\sigma^2} \geq \frac{1}{2}, \\
0,&{\text{if}}& 0 \leq  \frac{P|h_{21}[t]|^2}{\sigma^2} < \frac{1}{2}.
\end{array}\right.
\end{equation}

\subsection{Corresponding Average Delay of ST}\label{sec:AESR_CAD}

Similar to the analysis in Subsection~\ref{sec:AAB_ADRF}, if $|h_{01}[t]|^2 \geq |h_{21}[t]|^2$, we gain an instantaneous delay of ST $l_{01}[t]$ through letting $\Gamma[t] = 1+\frac{P(|h_{01}[t]|^2-|h_{21}[t]|^2)}{\sigma^2+2P|h_{21}[t]|^2}$ and $\Delta\Big[t+l(t)\Big] = 1+\frac{(1-\eta[t+l(t)])P|h_{12}[t+l(t)]|^2}{\sigma^2}$ in (\ref{eq:ad}). Analogously, the instantaneous delay of ST $l_{21}[t]$ is also obtained when $|h_{01}[t]|^2 \leq |h_{21}[t]|^2$. The average delay of ST of the proposed suboptimal encoding/decoding solution can be written as ${\cal L}=\Big\{{\cal E}\{l_{01}[t]\},{\cal E}\{l_{21}[t]\}\Big\}$.

\section{Numerical Results}\label{sec:results}

Suppose that the channel gain $h_{ij}$, for $\{i,j\}\in\{0,1,2\}$, is modeled by a small-scale fading model with a distance path loss, given by $h_{ij}=\alpha_{ij} \cdot d_{ij}^{-\beta/2}$, where $\beta$ is the path loss exponent and fixed at 3, $d_{ij}$ and $\alpha_{ij}$ denote the distance between node $i$ and $j$ and the channel fading coefficient of the link from node $i$ to $j$ with ${\cal E}\{|\alpha_{ij}|^2\}=1$, respectively. Wherein, $\alpha_{01}$ and $\alpha_{21}$ follow Nakagami-m distribution and are i.i.d.. With $d_{ij}[t]=d_{ji}[t]$ and a constant $\beta$, $h_{01}$ and $h_{21}$ are also i.i.d.. As shown in Fig.~\ref{fig:system_model}, we assume that the distance between two source nodes is normalized to 1 and the location of the relay is determined using the projections $x$ and $y$. Two source nodes 0 and 2 are located at the coordinates (-0.5,0) and (0.5,0), respectively. Let $\{x,y \} \sim {\cal U}[-0.5,0.5]$, where ${\cal U}$ denotes Uniform distribution. Each node uses the same transmission power $P$ and the AWGN $z_j$ at node $j$ is subject to ${\cal CN}(0, \sigma^2)$.

\begin{figure}
\centering
\includegraphics[width=3.5in]{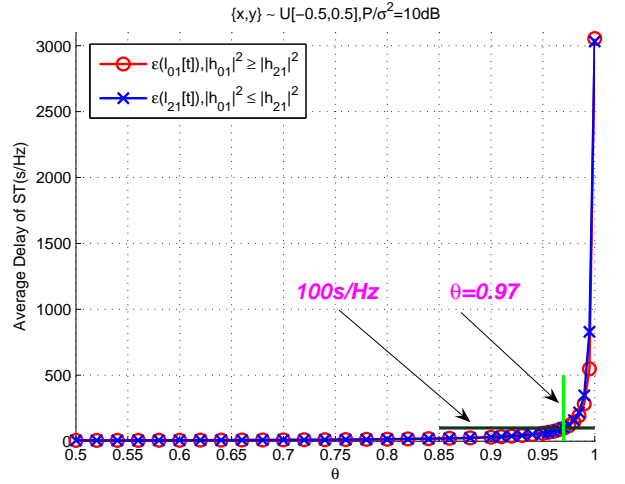}
\caption{Convergence behavior of average delay of ST for AAB protocol (Eq.~(\ref{eq:ad})).}
 \label{fig:delay_versus_theta}
\end{figure}

Fig.~\ref{fig:delay_versus_theta} illustrate the convergence behavior of the average delay of ST for the proposed AAB protocol. The correlation between the average delay of ST and $\theta$ is similar to an exponential function. The average delay of ST increases laxly and is less than about $100$s/Hz when $\theta < 0.97$ while grows sharply in the high $\theta$ regime. It is expected because the accumulative packets in the relay buffer can not be broadcasted to two source nodes in a limited time if $\theta < 0.97$. In other words, our proposed AAB protocol is appropriate for all TWR transmission cases, specially it can be used with a bounded average delay of ST (less than about 100 s/Hz) when $\theta < 0.97$.

\begin{figure}
\centering
\includegraphics[width=3.5in]{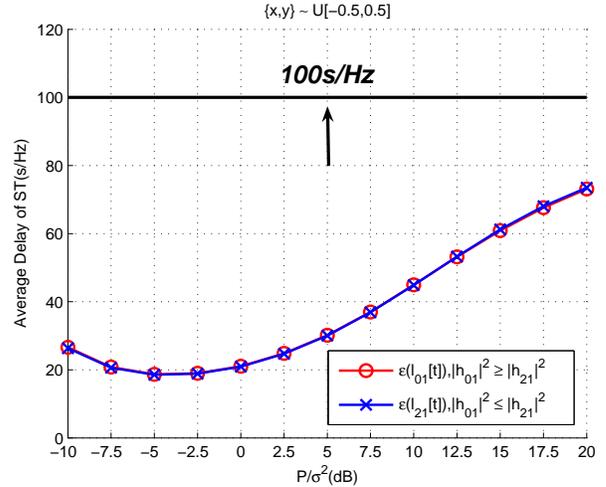}
\caption{Average delay of ST versus $P/\sigma^2$ for AAB protocol with a suboptimal encoding/decoding solution (Eq.~(\ref{eq:ad}) modified according to Subsection \ref{sec:AAB_ADRF}).}
 \label{fig:delay_versus_snr_suboptimal}
\end{figure}

Fig.~\ref{fig:delay_versus_snr_suboptimal} shows the variations of the average delay of ST for the AAB protocol with a suboptimal encoding/decoding solution when $P/\sigma^2$ is increasing. It can be clearly seen that the average delay of ST raises slowly as $P/\sigma^2$ is increasing and is always less than $100$s/Hz for all considered $P/\sigma^2$.
Fig.~\ref{fig:delay_versus_snr_suboptimal} confirms that our proposed AAB protocol is well realized by the proposed suboptimal encoding/decoding solution with a sufferable average delay of ST, even if we can not achieve the capacity of the TWR channels.

\begin{figure}
\centering
\includegraphics[width=3.5in]{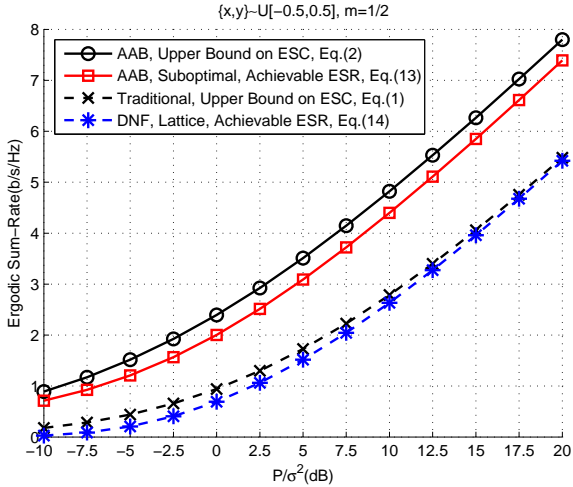}
\caption{Ergodic sum-rate (ESR) versus $P/\sigma^2$ in TWR Nakagami-m fading channels.}
 \label{fig:sumrate_versus_snr_m_12}
\end{figure}

Fig.~\ref{fig:sumrate_versus_snr_m_12} shows the ESR of different TWR protocols versus $P/\sigma^2$ in the TWR Nakagami-m fading channels. The upper bound on ESC of the proposed AAB protocol outperforms that of the traditional protocols about $2.4$b/s/Hz when $P/\sigma^2$ at $20$dB. The achievable ESR of AAB is only inferior to the upper bound with AAB about $0.4$b/s/Hz and is superior to the DNF protocol with Lattice code about $2$b/s/Hz at $P/\sigma^2=20$dB. The gaps between the achievable ESR of AAB and the upper bound with AAB are decreased slowly when $P/\sigma^2$ is increasing. At the same time, all the gaps between the achievable ESR of AAB and that of the traditional protocols are enlarged because of the influence of inferior channel gains.

\begin{figure}
\centering
\includegraphics[width=3.5in]{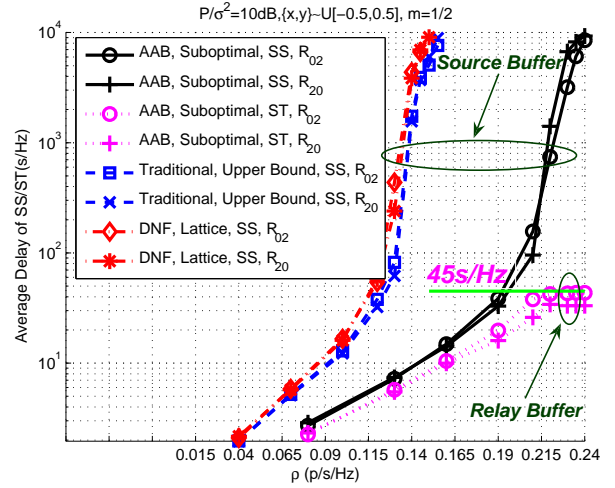}
\caption{Average delay of SS/ST versus packet arrival rate in TWR Nakagami-m fading channels.}
 \label{fig:delay_versus_queue_m_12}
\end{figure}

Suppose that both two source nodes have buffers for all considered TWR protocols while the relay node has buffer only for the proposed AAB protocol. We assume that the packet arrival rate (PAR), scaled by $p/s/Hz$, at two source nodes follows Poisson distribution with mean $\rho$, the length of each packet is fixed as 10 bits. Fig.~\ref{fig:delay_versus_queue_m_12} shows the average delay of SS/ST in the TWR Nakagami-m fading channels. The AAB protocol always outperforms the traditional protocols significantly. It is because that the AAB protocol can cancel the $\min$ operations between the uplink and downlink while the traditional protocols can not. For all considered TWR protocols, both the average delay of SS for two transmission directions in the source buffers are exponentially increased to about $10^4$s/Hz as the PARs are approaching the corresponding maximum values. Meanwhile, the average delay of ST in the relay buffer however stops increasing and maintains at about $45$s/Hz. Compared with the average delay of SS, the delay of ST induced by the AAB protocol is very small and negligible. In other words, the AAB protocol significantly improves the rate of information exchange with a sufferable delay of ST.

\section{Conclusion}

In this research, we proposed a new TWR protocol, named as alternative awaiting and broadcast (AAB).
The operational principles, new upper bound on ESC and convergence behavior of average delay of ST (in relay buffer) for the proposed AAB protocol are analyzed. The $\min$ operations between the uplink and downlink in an identical direction are canceled. We further derive an achievable ESR and the corresponding average delay of ST for the AAB protocol through presenting a suboptimal encoding/decoding solution. Numerical results show that 1) the proposed AAB protocol significantly improves the achievable ESR compared to the traditional TWR protocols, 2) considering the average delay of system service (SS) (in source buffer), the average delay of ST induced by the proposed AAB protocol is very small and negligible.

\bibliographystyle{IEEEtran}
\bibliography{IEEEabrv,reference}

\newpage

%
%

\end{document}